\documentclass{iaus}
\usepackage{graphicx}
\usepackage{natbib}

\newcommand{\be}{\begin{equation}}
\newcommand{\ee}{\end{equation}}
\newcommand{\bel}[1]{\begin{equation}\label{#1}}
\newcommand{\ba}{\begin{eqnarray}}
\newcommand{\ea}{\end{eqnarray}}
\newcommand{\bal}[1]{\begin{eqnarray}\label{#1}} 
\newcommand{\Odds}{\mathcal{O}}
\newcommand{\Bayes}{\mathcal{B}}
\newcommand{\pp}[2]{ \frac{ p(\textrm{\footnotesize{#1}}) }{ p(\textrm{\footnotesize{#2}}) } }
\newcommand{\snr}{\textrm{\footnotesize{SNR}}}
\newcommand{\mty}[1]{\textrm{\tiny{#1}}}

\title[Towards multimessenger astronomy] 
{Towards improving the prospects for coordinated gravitational-wave and electromagnetic observations}

\author[Ilya Mandel, Luke Z.~Kelley \& Enrico Ramirez-Ruiz]   
{Ilya Mandel$^1$, Luke Z.~Kelley$^2$
 \and Enrico Ramirez-Ruiz$^2$}

\affiliation{$^1$School of Physics and Astronomy, University of Birmingham, Edgbaston, 
B15 2TT, UK \\ email: {\tt imandel@star.sr.bham.ac.uk} \\[\affilskip]
$^2$ Harvard-Smithsonian Center for Astrophysics, 60 Garden Street, Cambridge, MA 02138
\\[\affilskip]
$^3$ Department of Astronomy and Astrophysics, University
  of California, Santa Cruz, CA 95064
  }

\pubyear{2011}
\volume{285}  
\pagerange{1--3}
\jname{New Horizons in Time-domain Astronomy}
\editors{E. Griffin, R. Hanisch, \& R. Seaman, eds.}

\begin{document}

\maketitle

\begin{abstract}
We discuss two approaches to searches for gravitational-wave (GW) and electromagnetic (EM) counterparts of binary neutron star mergers.  The first approach relies on triggering archival searches of GW detector data based on detections of EM transients.  We introduce a quantitative approach to evaluate the improvement to GW detector reach due to the extra information gained from the EM transient and the increased confidence in the presence of a signal from a binary merger.  We also advocate utilizing other transients in addition to short gamma ray bursts.  The second approach involves following up GW candidates with targeted EM observations.  We argue for the use of slower but optimal parameter-estimation techniques to localize the source on the sky, and for a more sophisticated use of astrophysical prior information, including galaxy catalogs, to find preferred followup locations. 
\end{abstract}

\firstsection 

\section{Introduction}

Advanced ground-based gravitational-wave detectors LIGO [\cite{AdvLIGO}] and Virgo [\cite{AdvVirgo}] are expected to begin taking data around 2015, operating at a sensitivity approximately a factor of 10 better than their initial counterparts.  Gravitational waves (GWs) emitted during the late inspirals and mergers of compact-object binaries will be one of the main sources for these detectors.  Although predictions for merger rates of binary neutron stars and neutron-star -- black-hole binaries are highly uncertain (see, e.g, \cite{MandelOshaughnessy:2010}), we may expect detections at a rate between once per few years and a few hundred per year, with perhaps a few tens of detections per year being most likely [\cite{ratesdoc}].  

These detections would usher in an era of genuine gravitational-wave astronomy, with GWs being used as another tool to observe the sky.  There has been a lot of discussion of the promise of multimessenger observations in the literature (see, e.g., \cite{Bloom:2009}).  In fact, the recent focus on multimessenger astronomy has been so exclusive that it is worthwhile to remind ourselves that a significant amount of astrophysics can be extracted from the GW observations alone, since the GW signal encodes the masses and spins of the binary components, and can be used to probe astrophysics, strong-field gravity, and cosmology even in the absence of electromagnetic (EM) observations of counterparts to GW events.  Nonetheless, there is no doubt that observing both EM and GW counterparts of the same event would be of great astrophysical significance, and would allow us to settle crucial questions such as the origins of short GRBs.

In general, such observations can be achieved in three different ways: (i) through entirely independent, serendipitous observations of GW and EM signatures of the same event, (ii) through triggered searches of archival GW data based on EM transients observed during surveys, or (iii) through telescope pointing to search for electromagnetic followups of GW candidates.  Here, we describe some thoughts about future possibilities for triggered GW searches based on observed EM transients, and possible improvements to recently started efforts to follow up GW triggers with targeted EM observations. 

\vspace{-0.15in}
\section{GW searches triggered on EM transients}

The detection of an electromagnetic transient which may originate from a compact-object binary merger will increase the a priori probability that a given stretch of data from the LIGO-Virgo ground-based gravitational-wave detector network contains a signal from a binary coalescence. ÊAdditional information contained in the electromagnetic signal, such as the sky location or distance to the source, can further help to rule out false alarms, and thus lower the necessary threshold for a detection.  

The LIGO Scientific Collaboration and the Virgo Collaboration have previously used short, hard Gamma Ray Bursts (SGRBs) as triggers for searches for compact binary coalescences in GW detector data [\cite{S5GRBLV}].  SGRBs are believed to be associated with relativistic jets from BNS or NS-BH mergers (see, e.g., \cite{LeeRamirezRuiz:2007}).  However, the impact of the SGRB observation on gaining confidence in the presence of a GW signature from a binary merger has not been quantitatively estimated.   In fact, most of the observed SGRBs originate too far away to detect the associated GW signal: the average luminosity distance for the 16 SGRBs with confident host identifications and redshift measurements compiled by \cite{Berger:2010} is approximately 5 Gpc, which exceeds the anticipated event horizon of Advanced LIGO by more than a factor of 10.  Hence, even when an SGRB is detected, it is unlikely that a gravitational-wave counterpart is observable, so the detectability threshold is not lowered by as much as could be anticipated.

\cite{Kelley:2011} develop a Bayesian framework to accurately incorporate the additional prior information on the existence of a signal and its timing, sky location, and potentially distance and inclination, in order to quantify the benefits of triggered searches in improving the detector reach.    The odds ratio for the GW detector data $d$ to contain a signal relative to the noise-only model is 
	\be \label{eq_odds}
	\Odds \equiv \pp{GW$|$d}{N$|$d} = \pp{GW}{N} \pp{d$|$GW}{d$|$N} ,
	\ee
where the first term is the a priori probability of having a GW signal present, and the second term is known as the Bayes factor,
\be
\Bayes = \int  p(\theta|GW) \cdot e^{<d|h(\theta)> - \frac{1}{2}<h(\theta)|h(\theta)>} d\theta ,
\ee
where the evidence integral is computed over all possible values of the parameters $\theta$ of the signal model with their corresponding priors $p(\theta|\textrm{GW})$.  We can approximate the Bayes factor by assuming that the likelihood in the previous integral is roughly constant over a fraction $\eta$ of the parameter space for which the signal model is supported by the data, where it has a simple dependence on the signal-to-noise ratio (SNR), and zero elsewhere:
	\be  \label{eq_odds_snr}
	\Odds \propto \pp{GW}{N} \cdot \eta \cdot e^{ \frac{1}{2} ( \snr )^2 } .
	\ee
	
The first term in Eq.~(\ref{eq_odds_snr}) can be increased by the detection of an EM transient, depending on our a priori confidence that this kind of transient is associated with a binary coalescence, the probability that the coalescence occurs within the detection volume, and the accuracy of the merger time reconstruction from the transient.  The second term is also affected by the additional information from the transient on the sky location of the source, as well as possibly the distance to the source and its inclination, which constrains the priors on individual parameters and increase the fraction of the parameter space which offers support for the signal model.  Assuming a fixed threshold on the odds ratio required for detection (corresponding to a fixed false alarm probability), the EM transient can therefore reduce the SNR threshold for detection by the following fraction:
	\be  \label{eq_snr_ratio}
	\zeta_\snr \equiv \frac{ \snr_\mty{EM} }{ \snr } = 	\left[  \frac{ \textrm{ ln}\left( \Odds_\mty{EM} \cdot \left[ \pp{GW$|$EM}{N$|$EM} \cdot \eta_\mty{EM} \right]^{-1} \right) }
								        { \textrm{ ln}\left( \Odds \cdot \left[ \pp{GW}{N} \cdot \eta \right]^{-1} \right) }
							\right]^\frac{1}{2}.
	\ee

Preliminary results [\cite{Kelley:2011}] suggest that optical signatures from r-processes in the tidal tails (also known as "kilonovae") could be one of the most promising triggers. They are likely to be sufficiently numerous in all-sky surveys like the LSST, differentiable from other transients based on the light curve profiles if the observational cadence is sufficiently high, and could yield moderate improvements in the detectability thresholds.

The above formalism assumes that optimal coherent searches are used both in the presence and in the absence of an EM trigger.  If the untriggered search is a suboptimal incoherent search while the triggered search is optimized because of the known sky location [\cite{HarryFairhurst:2011}], the improvement can be even more significant.

\vspace{-0.15in}
\section{EM followups of GW candidate events}

The benefits of following up GW candidates with targeted EM observations have long been recognized [\cite{Finn:1999}].  Recently, the first pilot program for  following up GW candidate events has been activated [\cite{S6BurstFollowup}].  There are two significant complications in this effort.  One is the speed with which GW data can be analyzed to yield information about detection confidence and the sky location of GW candidates.  The other is the large positional uncertainty associated with GW detections, which could encompass tens or even a few hundred square degrees on the sky depending on the SNR of the candidate and the detector network configuration (e.g., \cite{Nissanke:2011}).  

A significant amount of effort has been expended to allow for very rapid processing of GW data with the aim of achieving latencies of only a few seconds or tens of seconds to identify GW candidates (e.g., \cite{LLOID}).  Timing triangulation between different GW interferometers comprising the network has been used so far to rapidly localize the source on the sky (e.g., \cite{Fairhurst:2009}).  Such incoherent use of detector data allows for rapid analysis, but is likely to yield suboptimal results.  The ``need for speed'' can be over-stated: it is simply not necessary for many of the possible EM counterparts.  For example, off-axis optical afterglows outside the jet opening angle will only peak on time scales greater than $\sim 1$ day [\cite{vanEertenMacFadyen:2011}].  We may hope that if we are inside the jet opening angle, the GRB will be picked up by a mission like Swift; but even if not, the optical signal has decay timescales of at least several hours.  Timescales for lower-frequency followups, such as radio, are even longer.  We thus have the luxury of applying slower parameter-estimation techniques that perform a coherent analysis of all of the detector data (e.g., \cite{vanderSluys:2008b}), and should be able to improve sky localization and better quantify the confidence regions [\cite{Aylott:2012}].

Even with slower, optimal data-analysis techniques, the confidence regions are much larger than the fields of view of typical followup instruments.  Efforts to reduce the GW sky error box have relied on using a galaxy catalog [\cite{NuttallSutton:2010}].  In particular, pixels within the error box have been re-weighted by the blue-light luminosity of the galaxies contained in them [\cite{S6BurstFollowup}].   While a reasonable first cut, this approach could be problematic for three reasons.  Firstly, blue-light luminosity is a proxy for the star formation rate, but mergers could be significantly delayed relative to star formation episodes, and red elliptical galaxies are thought to contribute significantly to present-day merger rates [\cite{OShaughnessy:2009}].  Secondly, galaxy catalogs are unlikely to be complete to $\sim 400$ Mpc (the horizon distance of Advanced LIGO for optimally located and oriented coalescing neutron-star binaries).  Finally, some fraction of mergers may happen outside galaxies altogether, as the progenitor binary could have been kicked out of the hosts by supernovae kicks [\cite{Kelley:2010}].  An ongoing study by \cite{Vousden:2012} aims to account for these shortcomings by employing more sophisticated astrophysical priors.

Despite the improvements mentioned above, finding an electromagnetic counterpart of a GW candidate will remain extremely challenging, as discussed by \cite{MetzgerBerger:2011}.  Coordinated observing among several facilities may be required to cover the large uncertainty region with smaller field-of-view instruments.  Another alternative that may be worth investigating is the deployment of a network of inexpensive robotic telescopes specifically with the goal of following up GW candidates, though it will be difficult to detect any but the closest afterglows. While it is impossible to guarantee that an EM counterpart to given candidate would be found, we can still strive to maximize the probability of a successful follow up, with a view to ensuring that a sufficient fraction of triggers are followed up to make at least some multimessenger observations.

\bibliographystyle{apj}
\bibliography{Mandel.bib}

\end{document}